\documentclass[iop,apjl]{emulateapj}

\begin{document}

\newcommand{\grb}{GRB~130606A}
\newcommand{\lya}{Ly$\alpha$}
\newcommand{\lyb}{Ly$\beta$}
\newcommand{\lyg}{Ly$\gamma$}
\newcommand{\taueff}{\ensuremath{\tau_{\mathrm{GP}}^{\mathrm{eff}}}}
\newcommand{\ew}{\ensuremath{EW_{\mathrm{r}}}}
\newcommand{\lnh}{\ensuremath{\log(N_{\mathrm{H I}})}}

\shorttitle{GRB 130606A at $z=5.913$}
\shortauthors{Chornock et al.}

\title{GRB 130606A as a Probe of the Intergalactic Medium and the
  Interstellar Medium in a Star-forming Galaxy in the First Gyr
  After the Big Bang}

\author{Ryan Chornock\altaffilmark{1},
Edo Berger\altaffilmark{1},
Derek B. Fox\altaffilmark{2},
Ragnhild Lunnan\altaffilmark{1},
Maria R. Drout\altaffilmark{1},
Wen-fai Fong\altaffilmark{1},
Tanmoy Laskar\altaffilmark{1},
and Katherine C. Roth\altaffilmark{3}
}

\altaffiltext{1}{Harvard-Smithsonian Center for Astrophysics,
                 60 Garden Street, Cambridge, MA 02138, USA; 
                 \texttt{rchornock@cfa.harvard.edu}}
\altaffiltext{2}{Department of Astronomy and Astrophysics,
  Pennsylvania State University, 525 Davey Laboratory, University
  Park, PA 16802, USA}
\altaffiltext{3}{Gemini Observatory, 670 North Aohoku Place, Hilo, HI
  96720, USA}

\begin{abstract}
We present high signal-to-noise ratio Gemini and MMT spectroscopy of
the optical afterglow of the gamma-ray burst (GRB) 130606A at redshift
$z$=5.913, discovered by {\it Swift}.  This is the first
high-redshift GRB afterglow to have spectra of comparable quality to
those of $z\approx6$ quasars.  The data exhibit a
smooth continuum at near-infrared wavelengths that is sharply cut
off blueward of 8410~\AA\ due to absorption from \lya\ at redshift
$z\approx5.91$, with some flux transmitted through the \lya\ forest
between 7000$-$7800~\AA. We use column densities inferred from metal
absorption lines to constrain the metallicity of the host galaxy
between a lower limit of [Si/H]$\gtrsim$$-1.7$ and an upper limit of 
[S/H]$\lesssim$$-0.5$ set by the non-detection of \ion{S}{2}
absorption.  We demonstrate consistency between the dramatic
evolution in the transmission fraction of \lya\ seen in this spectrum
over the redshift range $z=4.9$ to 5.85 with that previously measured
from observations of high-redshift quasars.  There is an extended
redshift interval of $\Delta z$=0.12 in the \lya\ forest at $z$=5.77
with no detected transmission, leading to
a 3$\sigma$ upper limit on the mean \lya\ transmission fraction
of $\lesssim$0.2\% (or \taueff(\lya)$>$6.4).  This is comparable to
the lowest-redshift Gunn-Peterson troughs found in quasar spectra.
Some \lyb\ and \lyg\ transmission is detected in this redshift window, 
indicating that it is not completely opaque, and hence that the IGM is
nonetheless mostly ionized at these redshifts.  We set a 2$\sigma$
upper limit of 0.11 on the neutral fraction of the IGM at the redshift
of the GRB from the lack of a \lya\ red damping wing, assuming a
model with a constant neutral density.
\grb\ thus for the first time realizes the promise of GRBs as probes
of the first galaxies and cosmic reionization.
\end{abstract}

\section{Introduction}

Observations of Gunn-Peterson absorption troughs \citep{gp65} detected
in the spectra of quasars at redshifts $z\approx6$
\citep{becker01} have been interpreted as representing
the end stages of the reionization of the intergalactic medium (IGM;
e.g., \citealt{fan06}).  However, the polarization of
the cosmic microwave background radiation implies a higher typical
redshift for reionization \citep{wmap}.  These observations indicate
that reionization was likely a complex process that occurred over a
range in cosmic times with strong local variations.  Long-duration
gamma-ray bursts (GRBs) are produced by the deaths of massive stars
(e.g., \citealt{wb}) and offer the promise of being important probes
of this process with their highly luminous afterglows being detectable
to large redshifts \citep{lamb00}.

The most useful probe of the IGM opacity at high redshifts has been
\lya\ seen in absorption of quasars, but quasars have some
disadvantages as probes.  At low redshift, quasar measurements can
interpolate across absorptions in the \lya\ forest, but at higher
redshift the continuum is so highly absorbed that the proper level has
to be inferred.  The smooth power-law spectra of GRB afterglows are
intrinsically much simpler than the complicated spectra
of quasars, which have emission lines and sometimes broad absorption.  
Aside from practical matters, quasars may not be unbiased probes of
the high-redshift universe.  The ultraviolet (UV) emission from
quasars ionizes the region around them through the proximity effect.  In
addition, \citet{mes10} has argued that the highest-redshift quasars
are hosted by massive dark matter halos that are highly biased tracers
of the underlying matter distribution.  Even ignoring the ionizing
radiation from the quasar, he finds that typical high-redshift quasars
are located in regions that are overionized relative to the average
due to the associated large-scale structure. GRBs are
associated with sites of massive-star formation and will
be more widely distributed at high redshift in galaxies of lower
masses \citep{tanvir12} than the rare massive black holes needed to
power the most luminous quasars.

There are now only three spectroscopically-confirmed GRBs at $z>6$
despite active follow-up efforts of suspected high-$z$ bursts: GRB
050904 at $z$=6.295 \citep{kawai06,haislip06,totani06}, GRB 080913 at
$z$=6.733 \citep{greiner09,patel10}, and GRB 090423 at $z\approx8.2$
\citep{tanvir09,salva}.  In addition, GRB~090429B has a photometric
redshift of $\sim$9.4 \citep{nino_z9}.  These objects have proven the
existence of GRBs at these early epochs, and are beginning to
demonstrate the application of GRBs to studies of star formation
in the early universe \citep{tanvir12}.  However, despite their
promise as bright probes \citep{lamb00}, high-$z$ GRB afterglow
studies of the IGM to date \citep{totani06,gallerani,patel10} have
been hindered by the limited signal-to-noise ratio (S/N) of the
available spectra. This has now changed 
with the discovery of \grb\ and our follow-up observations.

The {\it Swift} Burst Alert Telescope (BAT) triggered on \grb\ on 2013
June 6 at 21:04:39 (all dates and times are UT; \citealt{gcn14781}).
The high-energy emission was extended, with a duration of $T_{90}$ =
277$\pm$19~s as seen by BAT \citep{gcn14819}, firmly establishing
\grb\ as a member of the long-duration population of GRBs.
Subsequent ground-based followup observations located an optical
transient (e.g., \citealt{gcn14782,gcn14783}) that was brighter in the
near-infrared (NIR; \citealt{gcn14784,gcn14785}).  Initial
spectroscopy from the Gran Telescopio Canarias \citep{gcn14790}
revealed that the afterglow redshift was $z\approx6.1$, which was
subsequently refined to $z=5.913$ by several groups, including ours
\citep{gcn14796,gcn14798,gcn14816}.

We present an analysis of the optical spectra of the
afterglow of \grb, the first high-redshift GRB spectra to be
comparable in quality to those of typical $z\approx6$ quasars.  In
section 2, we describe the data acquisition and reduction.  We analyze
the metal absorption lines from the  host galaxy and intervening IGM
systems in section 3, and set constraints on the abundances in the ISM
of the $z=5.913$ host galaxy.  In section 4, we measure the properties
of the \lya, \lyb, and \lyg\ absorption of the IGM and compare
to previous observations of high-redshift quasars.  We discuss the
implications in section 5.

\section{Observations}

We observed the afterglow of \grb\ starting at 04:04 on 2013
June 7 using the Blue Channel spectrograph \citep{bluechan} on the
6.5-m MMT.  We obtained a set of four 1200~s spectra as \grb\ rose
from airmass 1.22 to 1.08 with a midpoint time of 04:45 ($\Delta
t=7.68$~hr after the BAT trigger). The 832 lines/mm grating and LP530 
order-blocking filter were used to cover the range
7460$-$9360~\AA. Our 1\arcsec -wide slit gave a spectral resolution of
2.0~\AA\ full width at half-maximum (FWHM) and was oriented at the
parallactic angle \citep{alex82} to reduce effects of differential
atmospheric dispersion.  We  
acquired the source by taking advantage of the excellent pointing of
the MMT to offset from a nearby bright star to the coordinates
for the afterglow rapidly distributed by \citet{gcn14783}.  These
coordinates are somewhat offset from the precise radio position given
by \citet{gcn14817}, possibly indicating that the object was not
fully in the slit. 

We subsequently obtained four 1800~s observations of \grb\ using the
Gemini Multi-Object Spectrograph (GMOS; \citealt{hook04}) on the 8-m
Gemini-North telescope, with a midpoint of 10:17 on 2013 June 7
($\Delta t=13.1$~hr). The spectra were obtained in nod-and-shuffle 
mode \citep{glazebrook01} with the R400 grating and RG610
order-blocking filter.  We took advantage of the new red-sensitive
deep depletion detectors to use a grating setup with coverage longward
of 1$\mu$m.  The excellent seeing (0.5$-$0.7\arcsec) over the course of
observations allowed us to have a spectral resolution of
$\sim$5~\AA\ over the observed spectral range of 6200$-$10500~\AA.
The grating angle was adjusted by 50~\AA\ between the second and third
observations to fill in CCD chip gaps. The 1\arcsec -wide slit was
oriented at a position angle of 90$\degr$, but the airmass was low
($<1.1$).

We use IRAF\footnote{IRAF is
  distributed by the National Optical Astronomy Observatories,
    which are operated by the Association of Universities for Research
    in Astronomy, Inc., under cooperative agreement with the National
    Science Foundation.} to perform basic two-dimensional image
processing and extract the spectra after removal of cosmic
rays \citep{lacos}.  We apply flux calibrations and correct for
telluric absorption using our own IDL procedures.  Two
aspects of the GMOS data reduction require special attention.  The
first is that because Gemini does not generally obtain standard stars
at the time of observations, the variable atmospheric H$_2$O
absorption strength can lead to errors in the correction for
telluric absorption in the strong band near 9400~\AA.  We obtain
archival observations of the standard star BD$+$28~4211 and are
careful to scale the H$_2$O portion of the telluric correction
separately from the correction at the O$_2$ absorption bands.  

The second effect is that the Gemini data 
were taken in nod-and-shuffle mode.  We reduce the Gemini data with
two methods, once after applying the expected pairwise subtraction of
the data from the two nod positions and once ignoring the nod pairs and
treating each spectrum as a normal long-slit observation.  The first
method leads to better control of systematic sky subtraction errors
produced by flat fielding errors and bright night sky emission lines,
but comes at a cost of a factor of $\sqrt2$ increase in
the Poisson errors in night sky dominated portions of the spectrum.  
In both cases, we align and stack the two-dimensional frames taken
with the same grating tilt angle prior to spectral extraction.
The two reductions are highly consistent, so we splice
them together at 8780~\AA, using the second method at
shorter wavelengths to obtain the best S/N,
while the nod-and-shuffle reduction was used at longer wavelengths
where the systematic residuals from sky subtraction are otherwise
problematic.

\begin{figure}
\epsscale{1.2}
\plotone{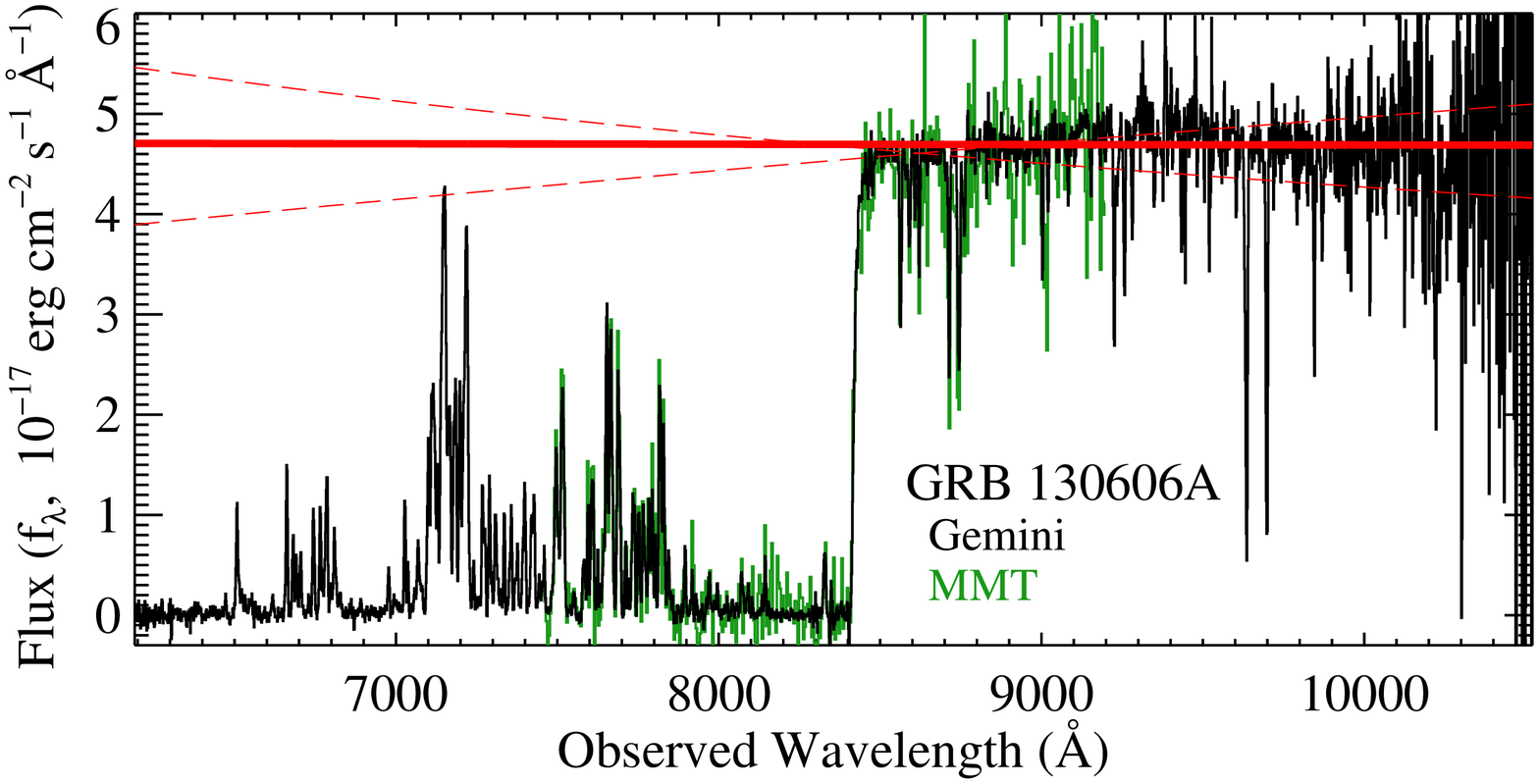}
\caption{Spectra of the optical afterglow of \grb\ from MMT/Blue
  Channel (green) and Gemini-N/GMOS (black), demonstrating impressive
  consistency after application of an arbitrary multiplicative scale
  factor.  The noisier MMT spectrum has been binned up by a
  factor of 6 for display purposes.  The solid red line shows
  a best-fit power law to line-free regions of the continuum.  The two
  dashed red lines show hypothetical extreme systematic variations
  around the best-fit value.  See Section 4.1 for details.
}
\label{specfig}
\end{figure}

We rebinned the calibrated one-dimensional spectra for each GMOS
grating setup to a common vacuum heliocentric wavelength scale and
combined them on a 
pixel-by-pixel basis, weighted by the inverse variance\footnote{We
  note that the sky is sufficiently bright that we are not in
  the regime pointed out by \citet{white03} where Poisson weighting
  biases the resulting flux level near zero counts.}.  The final MMT
and GMOS spectra are plotted in Figure~\ref{specfig}.  The absolute
flux scale is uncertain, both due to our use of an archival standard
star at Gemini and the fading of the afterglow that
is clearly evident during the observations.
Both spectra have been corrected for $E(B-V)=0.02$~mag of Galactic
extinction \citep{eddiedoug}.  In the continuum between 8500 and
8600~\AA, the MMT spectrum has a median S/N per 0.71~\AA\ pixel of
$\sim$10, while for the Gemini data the median S/N per 1.38~\AA\ pixel
is $\sim$80 and decreasing to longer wavelengths.

  Before proceeding further, we normalize
the spectra by fitting a power-law continuum to wavelength intervals
in the Gemini spectrum redward of the \lya\ break devoid of strong
absorption lines.  The best-fit continuum is shown as the thick red
line and has a slope of $f_{\lambda}$$\propto$$\lambda^{-0.01 \pm
  0.04}$, although some curvature relative to a pure power law is
evident.  We divide both spectra by this continuum in all subsequent
analysis.

\section{ISM Properties of a Star-forming Galaxy at $\lowercase{z}$=5.913}

The two spectra from different instruments presented in
Figure~\ref{specfig} are highly consistent with each other.  Both show a
flat continuum (in $f_{\lambda}$) at long wavelengths that drops
sharply down to zero between 8435 and 8405~\AA; flux detected
blueward of 8405~\AA\ is largely limited to wavelength intervals
between 6500 and 7850~\AA. This is indicative of absorption induced by
the host galaxy at redshift $z\approx5.91$ and the \lya\ forest at
lower redshift.

\begin{figure*}
\epsscale{1.2}
\plotone{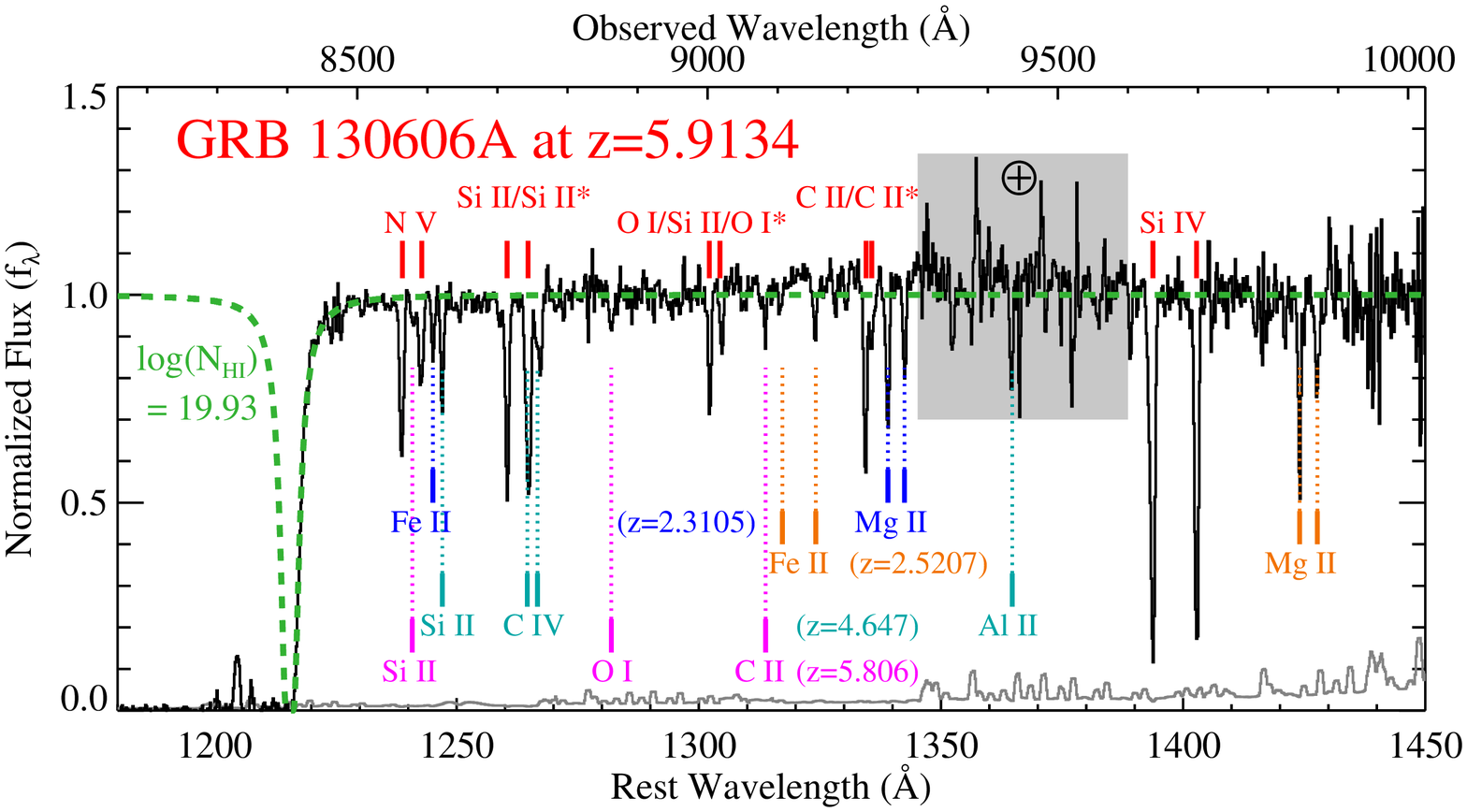}
\caption{GMOS spectrum of \grb\ with prominent absorption lines
  labeled.  Red lines and text identify absorption from the host
  galaxy at $z=5.913$. The green dashed line marks a fit to the
  \lya\ line at this redshift, with \lnh\ =19.93. The dramatic cutoff
  in flux at $\lambda$$<$1200~\AA\ caused by \lya\ at lower redshift is
  also apparent.  At least four intervening absorption systems are
  present at lower redshift and are labeled in different colors with
  their redshifts identified. 
  The gray box with the $\earth$ symbol marks the region possibly
  containing residuals from the correction for
  telluric H$_2$O absorption.  The gray line at the bottom is the
  error spectrum.
}
\label{linefig}
\end{figure*}

The normalized GMOS spectrum, displayed in Figure~\ref{linefig},
exhibits numerous absorption lines redward of \lya\ that arise in the
host galaxy and intervening systems.  We fit a local continuum and
single Gaussian profile to each absorption line in both the Gemini and
MMT 
spectra and list the results in Table~\ref{linetab}.  The quoted
uncertainties in line centroids and equivalent widths ($EW$) do not
include continuum placement uncertainties.  We are able to
confidently detect \ion{N}{5}, \ion{Si}{2}, \ion{Si}{2}*, \ion{O}{1},
\ion{O}{1}*, \ion{C}{2}, \ion{C}{2}*, and \ion{Si}{4} from the host
galaxy, as well as the red wing of \lya\ absorption at a similar
redshift.  A weighted average of the narrow, unblended, low-ionization
lines gives a redshift for \grb\ of $z=5.9134$, a value which we adopt
throughout this paper. The presence of fine-structure lines at this
redshift identifies it as that of the GRB because the lower levels for
these transitions are not normally populated unless pumped by the
UV emission from the GRB afterglow \citep{x_fine,vreeswijk07}.

In addition to absorption from the interstellar medium (ISM) of the
host galaxy, we detect absorption from at least four intervening
lower-redshift systems. \citet{gcn14816} previously reported two
absorption systems at $z=2.310$ and 3.451 in X-Shooter spectra.  We
confirm the existence of the first one through absorption from
\ion{Mg}{2} and \ion{Fe}{2} at an average redshift of
$z=2.3105\pm0.0001$.  We do not see the system at $z=3.451$, 
although no strong absorption lines from it are expected in our
observed wavelength range.  In addition, we detect at least three more
intervening systems, at weighted average redshifts of
$z=2.5207\pm0.0005$, 4.647$\pm$0.001, and 5.806$\pm$0.001.  The
$z$=5.806 \ion{O}{1}/\ion{C}{2}/\ion{Si}{2} system is particularly
interesting in light of recent observational work to find similar
systems at these redshifts to probe metal enrichment of the IGM
associated with reionization \citep{becker11}.  We note
that \ion{C}{4} $\lambda$1548 from the $z$=4.647 system completely
overlaps the \ion{Si}{2}* $\lambda$1265 absorption from the host and
that weak \ion{Fe}{2} $\lambda$2586 at $z$=2.3105 is also likely
blended with \ion{N}{5} $\lambda$1239.

\begin{figure}
\epsscale{1.2}
\plotone{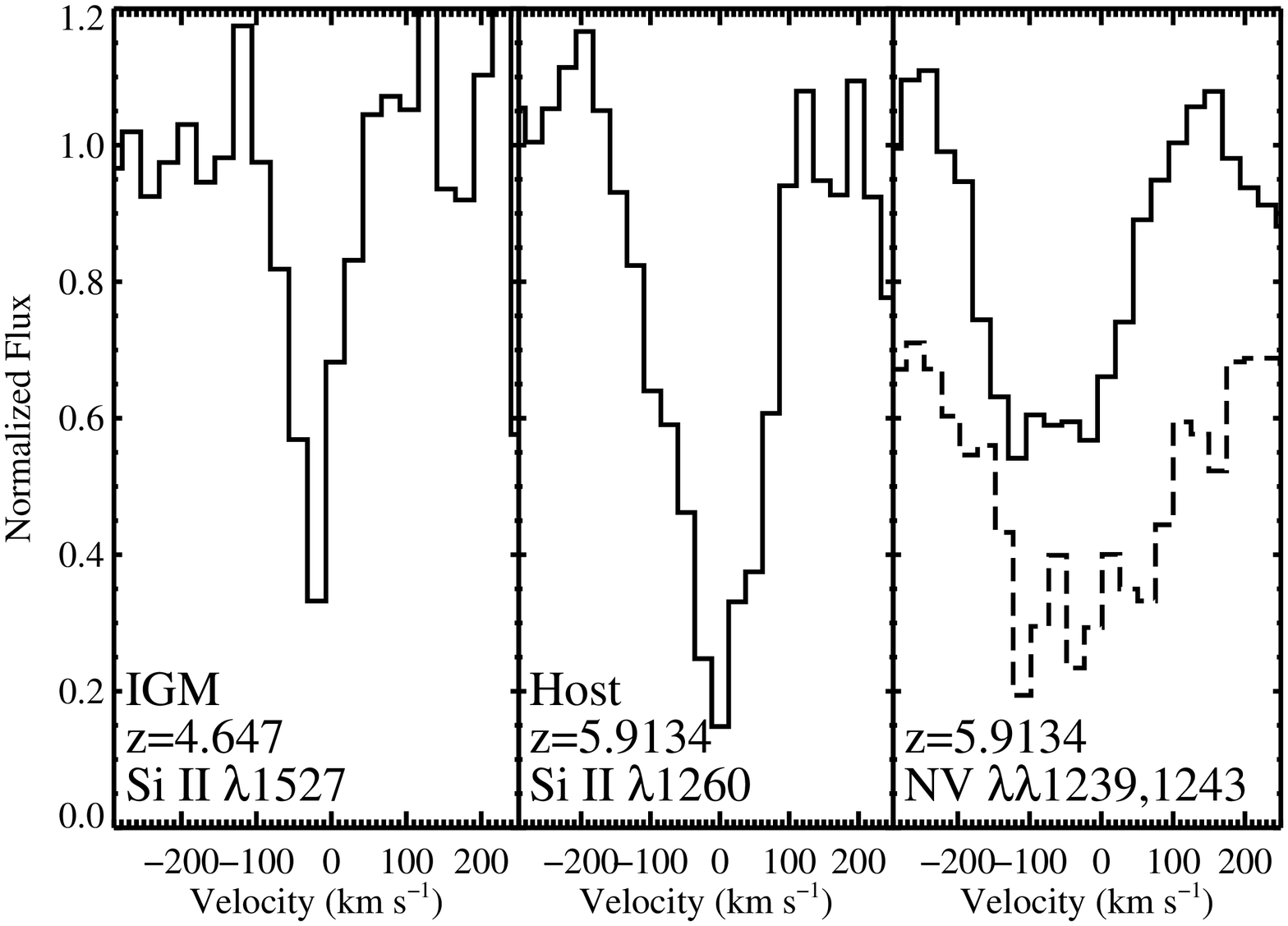}
\caption{Absorption line profiles in our MMT spectrum.  The left panel
  shows unresolved \ion{Si}{2} $\lambda$1527 absorption from a
  foreground absorber (FWHM$\approx$67 km~s$^{-1}$).  The middle panel
  shows the \ion{Si}{2} $\lambda$1260 absorption from the host, with
  intrinsic FWHM$\approx$120 km~s$^{-1}$.  The right panel shows the
  blueshifted and broader \ion{N}{5} doublet absorption from the host,
  with the $\lambda$1243 profile shifted downward by 0.4
  units. \ion{N}{5} $\lambda$1239 is possibly blended with \ion{Fe}{2}
  $\lambda$2586 from the $z$=2.3105 foreground absorber, but the
  profile is fairly consistent with the unblended $\lambda$1243.
}
\label{profilefig}
\end{figure}

One important advantage of the MMT data is that the host 
absorption lines are generally resolved, while the IGM metal lines are
not.   We show some unblended line profiles in Figure~\ref{profilefig}.
The \ion{Fe}{2} and \ion{Si}{2} lines from lower-redshift absorbers
have FWHMs consistent with the spectral resolution of $\sim$2~\AA.
However, the host \ion{Si}{2} $\lambda$1260 absorption has a FWHM
equivalent to $\sim$120 km~s$^{-1}$ after subtraction of the
instrumental width in quadrature.  Most of the host absorption lines
appear to be consistent with a single absorption component, but
the \ion{N}{5} lines are both blueshifted relative to the
low-ionization lines and have flat-bottomed profiles.
They exhibit absorption spread across $\sim$200
km~s$^{-1}$, which is unusually broad for \ion{N}{5} absorption in GRB
afterglows \citep{x_nv}.  The Gemini data 
cannot clearly resolve these features although some variation in FWHM is
apparent.  In particular, the \ion{Si}{4} doublet shows some
structure, with the stronger $\lambda$1394 line exhibiting a blue wing
extending out to a blueshift of $\sim$300 km~s$^{-1}$.  The \ion{N}{5}
and \ion{Si}{4} profiles probably
reflect absorption in a wind or outflow from the host galaxy. 

The sharp cutoff of flux between 1215 and 1220~\AA\ due to the wing of
\lya\ absorption in the host is indicative of a low hydrogen column
density (\lnh $<20.3$; all reported columns are in units of cm$^{-2}$)
in the host galaxy.  A fit to the hydrogen column in the GMOS
spectrum gives a best fit redshift of $z$=5.913$\pm$0.001, in excellent
agreement with the results from the narrow metal lines, and
\lnh=19.93$\pm$0.07, which we adopt in the subsequent analysis.  A fit
to the MMT spectrum with the redshift fixed gives a consistent value
of \lnh=19.99.  This value is rather low for a GRB host galaxy
\citep{palli06}, as it falls below the cutoff of \lnh=20.3 for a 
damped \lya\ system (DLA).
In the compilation of hydrogen column measurements of $z>4$ GRB host
galaxies presented by \citet{thoene2013}, only 1 out of 12 objects
(GRB 080913) has a lower \lnh\ of 19.84 \citep{greiner09,patel10}.

\begin{deluxetable*}{cccccc}
\tabletypesize{\scriptsize}
\tablecaption{Absorption Lines in \grb\ Spectra}
\tablehead{
\colhead{$\lambda_{\mathrm{obs}}$ (\AA)} &
\colhead{Line ID} &
\colhead{$\lambda_{\mathrm{rest}}$ (\AA)} &
\colhead{Redshift} &
\colhead{\ew\ (\AA)\tablenotemark{a}} &
\colhead{log($N_{\mathrm{X}}$, cm$^{-2}$)\tablenotemark{b}}
}
\startdata
\cutinhead{MMT spectrum}
8562.44$\pm$0.32 & \ion{N}{5} & 1238.82 & 5.9118$\pm$0.0003 & 0.37$\pm$0.06 & 14.24$\pm$0.07 \\
8590.87$\pm$0.41 & \ion{N}{5} & 1242.80 & 5.9125$\pm$0.0003 & 0.26$\pm$0.07 & 14.39$\pm$0.11 \\
8607.94$\pm$0.16 & \ion{Fe}{2} & 2600.17 & 2.3105$\pm$0.0001 & 0.21$\pm$0.07 & 13.16$\pm$0.14 \\
8620.67$\pm$0.19 & \ion{Si}{2} & 1526.71 & 4.6466$\pm$0.0001 & 0.23$\pm$0.07 & 13.94$\pm$0.13 \\
8703.00$\pm$0.42 & Unknown\tablenotemark{c} & \nodata & \nodata & 0.76$\pm$0.36 & \nodata \\
8713.65$\pm$0.15 & \ion{Si}{2} & 1260.42 & 5.9133$\pm$0.0001 & 0.50$\pm$0.06 & 13.55$\pm$0.05 \\
8744.28$\pm$0.32 & \ion{Si}{2}*\tablenotemark{d} & 1264.74 & 5.9139$\pm$0.0003 & 0.59$\pm$0.08 & 13.67$\pm$0.06 \\
\nodata & +\ion{C}{4}\tablenotemark{d} & 1548.20 & (4.647)  & & \\
8759.56$\pm$1.05 & \ion{C}{4} & 1550.77 & 4.6485$\pm$0.0007 & 0.36$\pm$0.13 & 14.25$\pm$0.16 \\
9002.73$\pm$0.34 & \ion{O}{1} & 1302.17 & 5.9136$\pm$0.0003 & 0.32$\pm$0.12 & 14.64$\pm$0.16 \\
\cutinhead{Gemini spectrum}
8563.33$\pm$0.05 & \ion{N}{5} & 1238.82 & 5.9125$\pm$0.0001 & 0.38$\pm$0.01 & 14.25$\pm$0.01 \\
8580.07$\pm$0.40 & \ion{Si}{2} & 1260.42 & 5.8073$\pm$0.0003 & 0.05$\pm$0.01 & 12.53$\pm$0.10 \\
8590.94$\pm$0.12 & \ion{N}{5} & 1242.80 & 5.9125$\pm$0.0001 & 0.21$\pm$0.01 & 14.30$\pm$0.02 \\
8608.02$\pm$0.13 & \ion{Fe}{2} & 2600.17 & 2.3106$\pm$0.0001 & 0.23$\pm$0.02 & 13.21$\pm$0.04 \\
8621.13$\pm$0.10 & \ion{Si}{2} & 1526.71 & 4.6469$\pm$0.0001 & 0.24$\pm$0.02 & 13.97$\pm$0.03 \\
\nodata & \ion{S}{2} & 1250.58 & (5.9134) & $\lesssim0.04$\tablenotemark{e} & $\lesssim14.7$ \\
\nodata & \ion{S}{2} & 1253.81 & (5.9134) & $\lesssim0.05$ & $\lesssim14.6$ \\
8703.51$\pm$0.40 & Unknown\tablenotemark{c} & \nodata & \nodata & 0.44$\pm$0.09 & \nodata \\
8714.09$\pm$0.04 & \ion{Si}{2} & 1260.42 & 5.9136$\pm$0.0001 & 0.43$\pm$0.01 & 13.48$\pm$0.01 \\
8744.31$\pm$0.05 & \ion{Si}{2}*\tablenotemark{d} & 1264.74 & 5.9139$\pm$0.0001 & 0.60$\pm$0.01 & 13.67$\pm$0.01 \\
\nodata & +\ion{C}{4}\tablenotemark{d} & 1548.20 & (4.647)  & & \\
8759.78$\pm$0.28 & \ion{C}{4} & 1550.77 & 4.6487$\pm$0.0002 & 0.29$\pm$0.03 & 14.15$\pm$0.04 \\
8862.76$\pm$0.47 & \ion{O}{1} & 1302.17 & 5.8062$\pm$0.0004 & 0.10$\pm$0.02 & 14.12$\pm$0.09 \\
9003.10$\pm$0.19 & \ion{O}{1} & 1302.17 & 5.9139$\pm$0.0001 & 0.22$\pm$0.03 & 14.47$\pm$0.05 \\
9020.18$\pm$0.31 & \ion{O}{1}*\tablenotemark{d} & 1304.86 & 5.9127$\pm$0.0002 & 0.12$\pm$0.02 & 14.25$\pm$0.08 \\
\nodata & +\ion{Si}{2}\tablenotemark{d} & 1304.37 & (5.9134) & & \\
9082.33$\pm$0.29 & \ion{C}{2} & 1334.53 & 5.8056$\pm$0.0002 & 0.08$\pm$0.02 & 13.62$\pm$0.10 \\
9103.74$\pm$0.64 & \ion{Fe}{2} & 2586.65 & 2.5195$\pm$0.0002 & 0.10$\pm$0.04 & 13.40$\pm$0.17 \\
9153.45$\pm$0.33 & \ion{Fe}{2} & 2600.17 & 2.5203$\pm$0.0001 & 0.24$\pm$0.04 & 13.23$\pm$0.07 \\
9225.28$\pm$0.11 & \ion{C}{2} & 1334.53 & 5.9127$\pm$0.0001 & 0.35$\pm$0.02 & 14.24$\pm$0.03 \\
9234.04$\pm$0.42 & \ion{C}{2}* & 1335.71 & 5.9132$\pm$0.0003 & 0.18$\pm$0.03 & 14.01$\pm$0.07 \\
9257.06$\pm$0.11 & \ion{Mg}{2} & 2796.35 & 2.3104$\pm$0.0001 & 0.66$\pm$0.04 & 13.19$\pm$0.03 \\
9281.16$\pm$0.16 & \ion{Mg}{2} & 2803.53 & 2.3105$\pm$0.0001 & 0.41$\pm$0.04 & 13.28$\pm$0.04 \\
9433.75$\pm$0.17 & \ion{Al}{2} & 1670.79 & 4.6463$\pm$0.0001 & 0.27$\pm$0.03 & 12.76$\pm$0.05 \\
9635.33$\pm$0.07 & \ion{Si}{4} & 1393.76 & 5.9132$\pm$0.0001 & 1.05$\pm$0.03 & 14.06$\pm$0.01 \\
9698.09$\pm$0.10 & \ion{Si}{4} & 1402.77 & 5.9135$\pm$0.0001 & 0.86$\pm$0.04 & 14.27$\pm$0.02 \\
9845.29$\pm$0.15 & \ion{Mg}{2} & 2796.35 & 2.5208$\pm$0.0001 & 0.65$\pm$0.06 & 13.18$\pm$0.04 \\
9870.25$\pm$0.62 & \ion{Mg}{2} & 2803.53 & 2.5206$\pm$0.0002 & 0.44$\pm$0.11 & 13.31$\pm$0.11 \\
\enddata
\tablenotetext{a}{Rest frame}
\tablenotetext{b}{Lower limit due to optically-thin assumption}
\tablenotetext{c}{$EW$ for unidentified lines are in observer frame}
\tablenotetext{d}{Redshifts and column densities are
  estimated assuming that the stronger component of blend dominates}
\tablenotetext{e}{Upper limits are 3$\sigma$}
\label{linetab}
\end{deluxetable*}

We set a lower limit to the metallicity by converting our absorption
line measurements into column densities of the metal ions.  In the
optically-thin limit, the column $N_{\mathrm X}$ of a given ion is
given by 
\begin{equation}
\mathrm{log}(N_{\mathrm X}) = 1.23 \times10^{20} \mathrm{cm}^{-2}
\frac{\ew({\mathrm \AA})}{\lambda_{\mathrm r} ({\mathrm \AA})^2  f_{ij}} ,
\end{equation}
where $\lambda_{\mathrm r}$ and \ew\ are the wavelength and equivalent
width of the transition in the rest frame, respectively, while
$f_{ij}$ is the oscillator strength.  We use the atomic data
collected by \citet{graasp}, and report the results in the
rightmost column of Table~\ref{linetab}.

The numbers we present are lower limits to the column density because
we make the assumption that all lines are optically thin.  This is
clearly wrong for the deep \ion{Si}{4} doublet and may be in error for
other transitions as well (e.g., the \ion{Si}{2} $\lambda$1260 profile
in Figure~\ref{profilefig}).  High-resolution spectroscopy of GRB
afterglows has revealed that the line profiles can have deep saturated
cores on top of absorption from lower columns.  These saturation
effects lead to systematic biases in metallicity measurements for GRB
hosts \citep{x_cog}, but the errors incurred are in the direction of
making 
the observed columns too low.  We can also take some confidence from
the fact that our inferred columns are completely consistent between
the GMOS data and the MMT data, which have a factor $\sim$3 higher
spectral resolution.  In fact, our MMT resolution is $R=\lambda
/\Delta\lambda \approx 4500$, approaching the moderate resolution of
spectrographs such as ESI at Keck or X-Shooter at VLT.  In addition,
ionization and dust depletion effects can cause us to underestimate
the column densities of some elements.

With those caveats in mind, we use the derived columns in
Table~\ref{linetab}, our value for \lnh\ from the fit to the wing of
\lya\ absorption, and the solar photospheric abundances of
\citet{asplund09} to constrain the metallicity of the host galaxy of
\grb.  High-ionization species such as \ion{Si}{4} and \ion{N}{5}
are not useful for a metallicity analysis as they trace more heavily
ionized gas that may not contribute to \lnh, as well as in this case
having clearly different velocity distributions from the narrow lines
(as described above).  This leaves \ion{O}{1}, \ion{C}{2}, and
\ion{Si}{2} as possible tracers, in addition to their excited
fine-structure transitions.  In the MMT data, we measure
[O/H]$\approx$$-2.0$ for \ion{O}{1} $\lambda$1302 only.  The Gemini
spectrum has a similar \ion{O}{1} column, but also exhibits
absorption\footnote{This line is blended with \ion{Si}{2} 
  $\lambda$1304, but we ascribe it completely  to \ion{O}{1}*, which
  is more consistent with the central wavelength of the absorption
  line.} from \ion{O}{1}*.
The combined oxygen abundance in those data is [O/H]$\approx$$-1.95$.
The formal uncertainties for these abundance measurements are
$\sim$0.1~dex, but the systematic errors clearly dominate.
Adding the \ion{C}{2} and \ion{C}{2}* columns from GMOS gives a
similar value of [C/H]$\approx$$-1.9$.  Determining the silicon column
is a little more challenging.  The foreground \ion{C}{4} $\lambda$1548
absorption 
coincides with the strongest \ion{Si}{2}* absorption line.  We correct
the \ion{Si}{2}* $\lambda$1265 observed-frame $EW$ value for
\ion{C}{4} by subtracting an observed-frame $EW$ of $\lambda$1548
equal to that measured for $\lambda$1551 and recalculate the column.
The exact correction factor has only a small effect on the final
result. Combining the \ion{Si}{2} and corrected \ion{Si}{2}* columns
gives a consistent estimate of [Si/H]$\approx$$-1.7$ in both the MMT
and Gemini spectra.

As described before, our measurements are all lower limits because of
dust depletion and the lack of constraints on other ionization stages
of the same elements.  However, we can also set an upper limit on the
metallicity from the non-detection of \ion{S}{2} lines.  Sulfur has
proven useful in previous studies of GRB host galaxies at high
redshift (e.g., \citealt{kawai06,berger06,price07}) because the weak
\ion{S}{2} lines near 1250~\AA\ are not likely to be saturated and
sulfur does not deplete onto dust grains \citep{savage}.  An
unidentified absorption line near 8703~\AA\ in the blue wing of
\ion{Si}{2} $\lambda$1260 is 
near the expected position of \ion{S}{2} $\lambda$1259.52 at the
redshift of \grb, but formally the centroid is more than
10$\sigma$ away from the correct wavelength, so we regard it as a
non-detection.  In addition, an implied \ion{S}{2} column that large 
should also produce $\lambda$1254 absorption that is not observed.  We
estimate observed-frame $EW$ 3$\sigma$ upper limits of
$\sim$0.3~\AA\ for any absorption line near \ion{S}{2}
$\lambda\lambda$1250, 1254.  These translate into an upper limit on
the metallicity of [S/H]$\lesssim$$-0.5$.

\begin{figure}
\epsscale{1.2}
\plotone{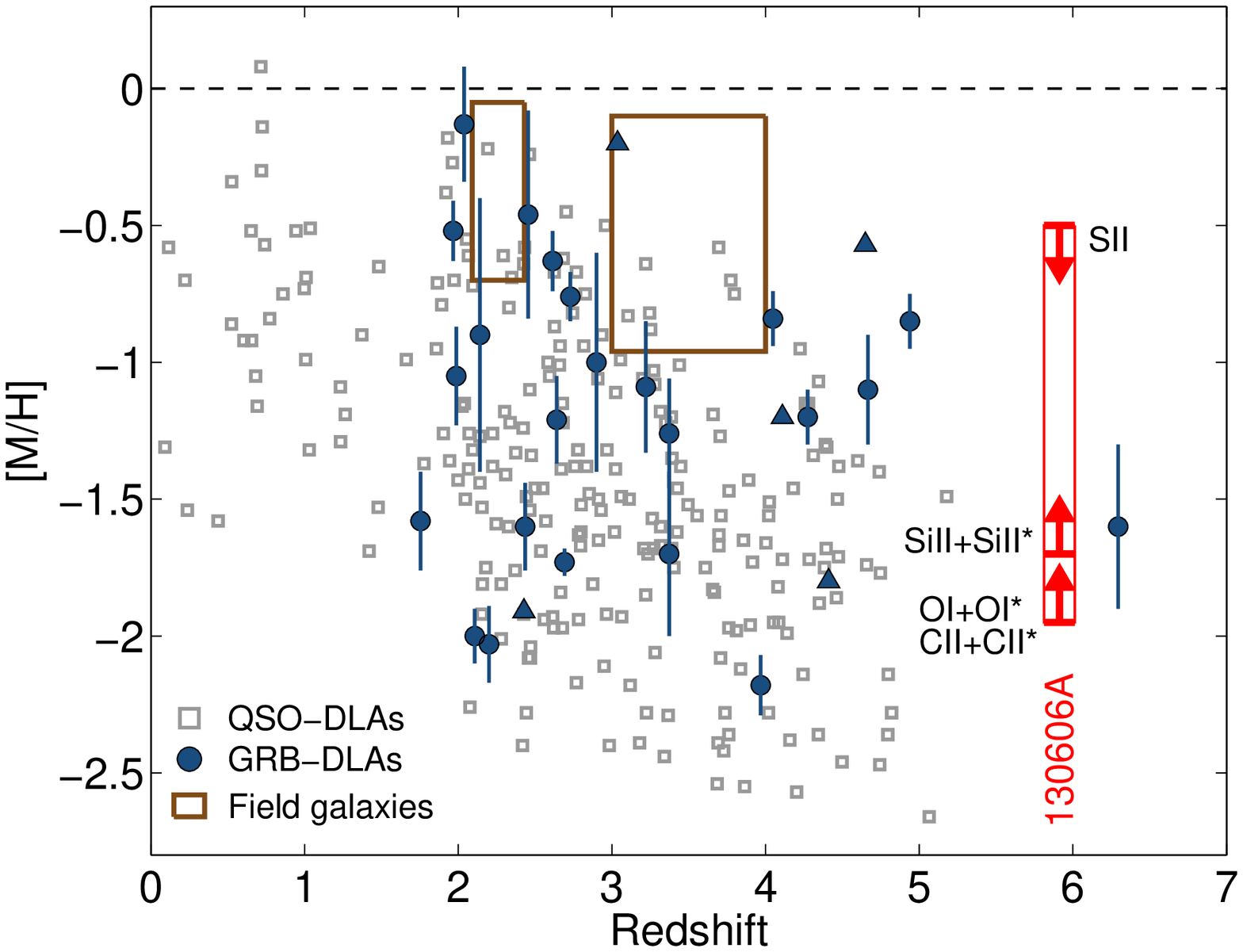}
\caption{Abundance estimates for the host of \grb\ compared to DLA
  systems in 
  both GRB host galaxies and quasars \citep{rafelski,thoene2013}.  The
  brown boxes represent the range in metallicity for star-forming
  10$^9-$10$^{11}$~M$_{\sun}$ galaxies at $z\approx2.3$ \citep{erb06}
  and $z\approx3.1$ \citep{mannucci09}.  These results unavoidably
  conflate measurements using a number of different tracers (e.g.,
  oxygen emission lines in the the field galaxies and metal absorption
  lines in the DLA samples).
}
\label{metalfig}
\end{figure}

We compare these metallicity contraints for \grb\ to abundance
measurements for DLA systems in both GRB host galaxies
\citep{thoene2013} and quasar absorption systems \citep{rafelski} in
Figure~\ref{metalfig}.  We also show the ranges reported in
star-forming field galaxy samples in the mass range of
10$^9-$10$^{11}$~M$_{\sun}$ at two different redshifts of
$z\approx2.3$ and $z\approx3.1$ \citep{erb06,mannucci09}.  The
metallicity range for \grb\ is at the low end of the dispersion in
field galaxy samples at lower redshift, but comparable to the GRB DLA
sample.

The point in Figure~\ref{metalfig} at $z=6.295$ from GRB~050904
\citep{kawai06} is of special interest because it represents the most
complete abundance analysis of a galaxy at a redshift comparable to
\grb.  \citet{kawai06} measured the abundance pattern using the exact
same lines we use here, except that they detect the \ion{S}{2}
lines, so systematic issues due to different tracers are minimized.
They found [C/H]$\approx$$-2.4$, [O/H]$\approx$$-2.3$, 
[Si/H]$\approx$$-2.6$, and [S/H]$\approx$$-1.0$. In comparison with our
results from above, the individual numbers are all $\sim$0.5~dex
higher in the host of \grb\ than in that of GRB~050904, likely
indicating a metallicity difference of about that magnitude with
a similar depletion pattern.  It is interesting to note that even at
these high redshifts, the ISMs of these star-forming galaxies show
clear evidence of chemical enrichment. 

\section{Opacity of the IGM at $\lowercase{z}\approx5-6$}

The spectra we present in Figures~\ref{specfig} and \ref{linefig}
exhibit a well-detected NIR continuum that drops to near zero at
8400~\AA, but then slowly rises to a peak near 7100~\AA\ before
turning over and dropping off blueward of that.  This continuum slope
in the absorbed part of the spectrum represents real evolution in the
optical depth of the \lya\ forest over the redshift range $5<z<6$. 
The very high S/N of our data allow us to place constraints on the
opacity of the IGM to \lya\ that are comparable to those from
individual high-redshift quasars.

\begin{figure*}
\epsscale{1.2}
\plotone{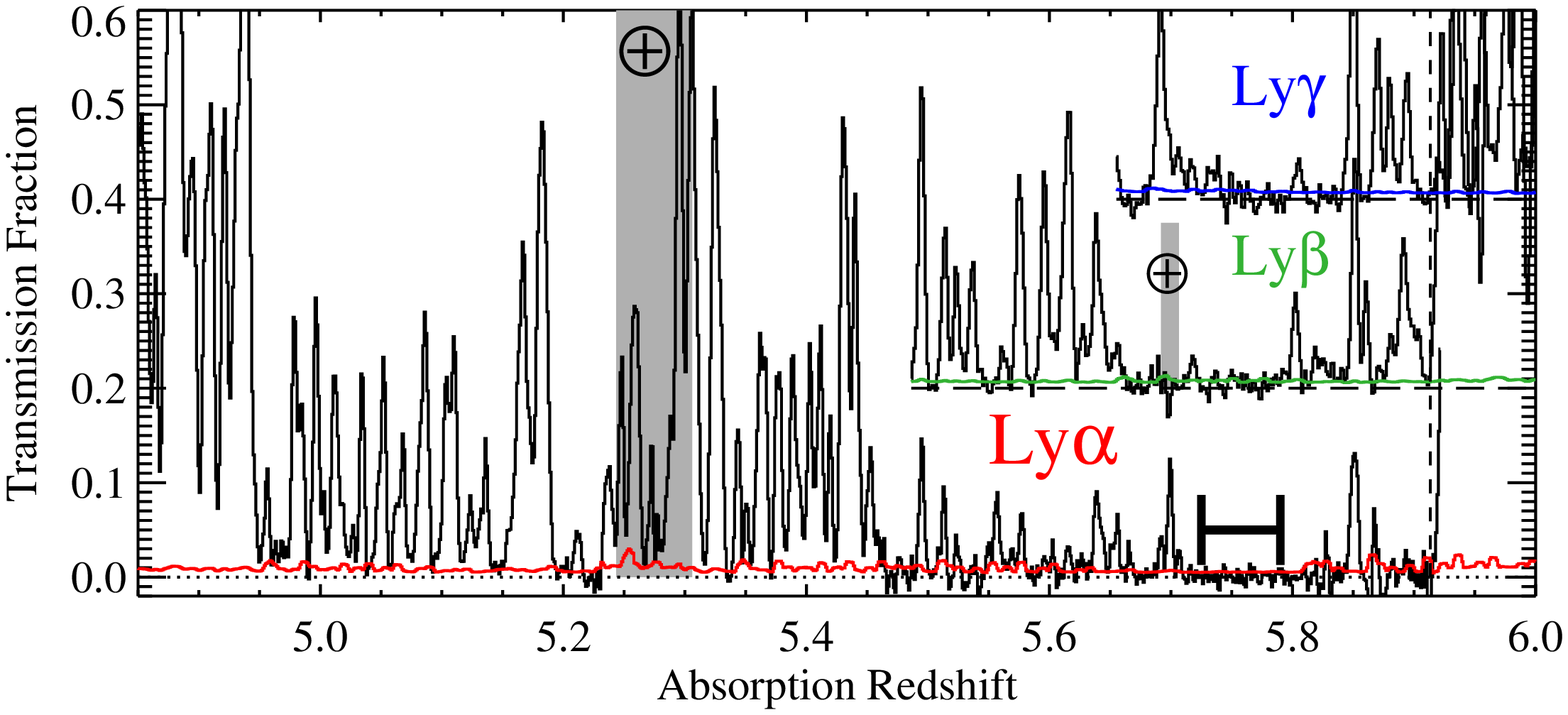}
\caption{Transmission fraction of Lyman-series transitions versus
  redshift. The spectrum corresponding to \lyb\ has been 
  shifted upward by 0.2 and that of \lyg\ by 0.4, with the horizontal
  dashed lines marking the new zero levels.  The red, green, and
  blue lines represent the 1$\sigma$ uncertainties about zero.
  The gray bars with $\earth$ symbols on the \lya\ and \lyb\ spectra
  represent regions possibly affected by imperfect correction for
  telluric absorption from the A and B bands, respectively.
  The black bar from $z = 5.725-5.79$ marks the longest 
  dark trough present in \lya.  The vertical dashed line at
  $z=5.9134$ marks the inferred redshift of the host galaxy from metal
  lines. 
}
\label{transfig}
\end{figure*}

We divide our GMOS spectrum by the best-fit continuum
marked on Figure~\ref{specfig} and then display the spectrum in
Figure~\ref{transfig} with the wavelength scale converted
to redshift relative to \lya, \lyb, and \lyg.  The transmitted flux is
clearly broken up into a ``picket fence'' of individual windows of
transmission through the \lya\ forest.  These windows are rare at
$z\approx5.8$, but become increasingly common at lower redshift until
at $z<5$ they start to overlap.  Figure~\ref{specfig} demonstrates the
consistency of these windows of transmission in two spectra of
different resolutions. 

Comparison of the three Lyman-series transitions shows that the
pattern of transmission windows is generally the same over the limited
overlap region, with the weaker higher-order lines having greater
transmission than \lya, as expected. There are two interesting
exceptions.  The first is that there is a 
weak window of transmission present near $z\approx5.803$ in both
\lyb\ and \lyg, but not \lya, in a redshift interval that is otherwise
fairly dark.  This is intriguingly close to the redshift of the
$z$=5.806 system that we detect in metal lines, indicating a moderate
local increase in ionization (but not enough for \lya\ to become
transparent) correlated with the same large scale structure
hosting the absorber.  Spatial correlations between metal enrichment
and ionization are predicted to be sensitive probes of the
reionization process and the pollution of the IGM by the earliest
galaxies \citep{oh02,fl03,becker11}.

Second, there is a clear transmission window present in both \lya\ and
\lyg\ between $z$=5.69$-$5.70 that is missing from \lyb.  Although
this region falls squarely in the atmospheric B band, the spectrum has
been corrected for telluric absorption and some flux might be
expected to be detected in data of this quality.  Instead, we note
that the \lya\ absorption associated with the $z=4.647$ system we detect
in metal lines would lie at exactly this redshift relative to \lyb.
This serves as a cautionary reminder that some absorption we attribute
to the Lyman series can be due to either metal lines or \lya\ at lower
redshift. 

\begin{figure}
\epsscale{1.2}
\plotone{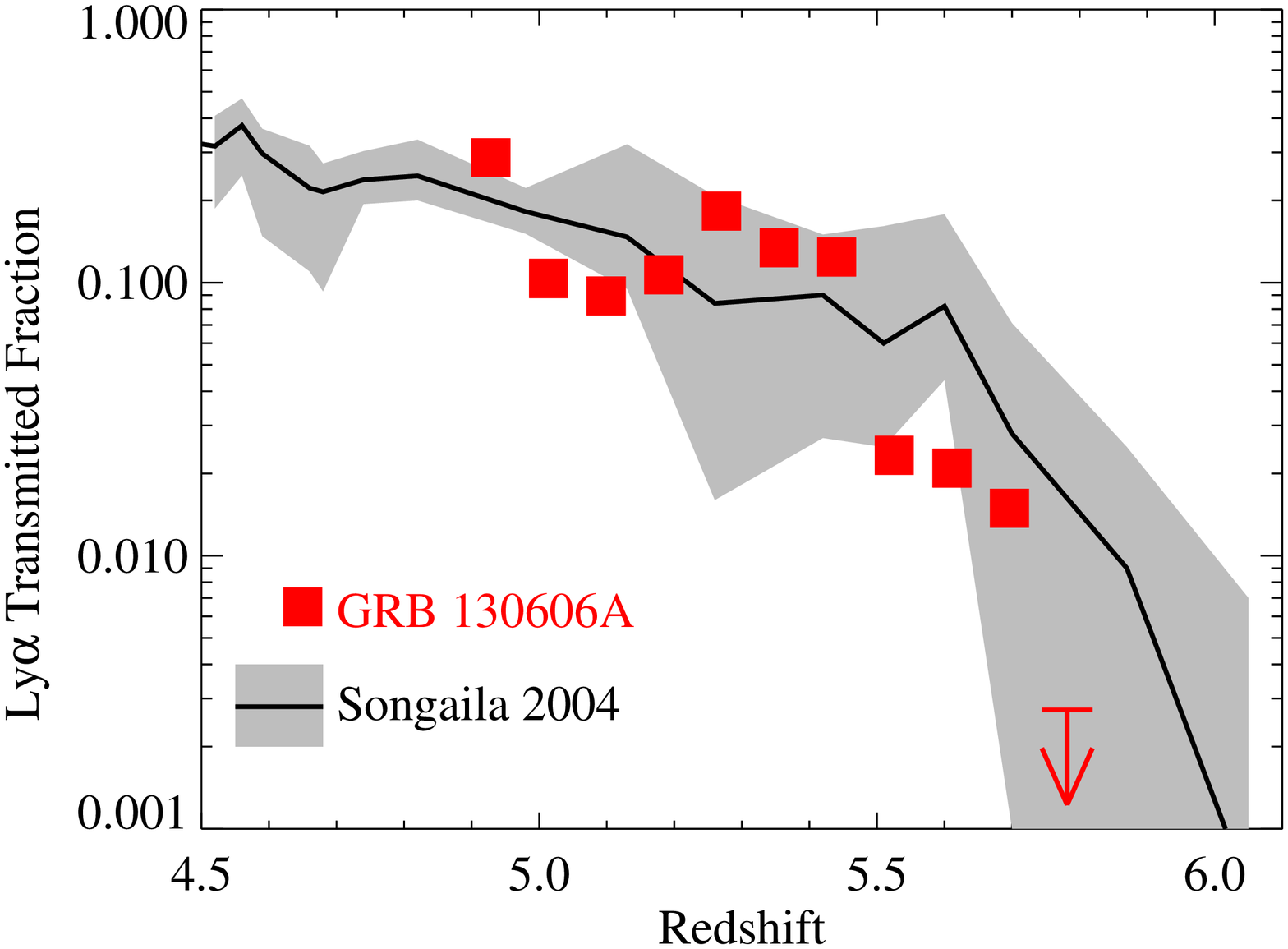}
\caption{Transmission fraction of \lya\ in 15~\AA\ bins from
  1035$-$1200~\AA\ (in the rest frame of \grb).  The thick black line
  marks the mean values determined by \citet{song04} with the gray
  region marking the range of observed values.  The plotted upper
  limit is 3$\sigma$.
}
\label{lyafig}
\end{figure}

We now compare the transmission along the line of sight to this
high-redshift GRB with previous studies using high-redshift
quasars as background light sources.  \citet{song04} used high-quality
moderate and high-resolution Keck spectra of a sample of quasars to
measure the mean transmission of \lya.  She computed the average
transmission in 15~\AA\ bins in the rest frame of each quasar between
1080 and 1185~\AA\ (limits chosen to avoid \lyb\ and \lya\ proximity
effects from each quasar) and
measured the mean and its variation along different lines of sight.
The thick solid line in Figure~\ref{lyafig} marks the mean
transmission measured from the quasars and the thick gray band marks
the observed minimum and maximum values in her sample.
We measure the transmission in wavelength bins of the same size over
the interval 1035$-$1200~\AA.  The low \lnh\ of the host of \grb\ 
allows us to measure the IGM opacity as close to the host
redshift as 1200~\AA\ without interference from the blue wing of \lya\
(cf. Figure~\ref{linefig}).  The red boxes in Figure~\ref{lyafig} mark
our measurements.  The formal error bars are far smaller than the
plotted symbols.  The strong fluctuations in our measurements above
and below the mean of the quasars are due to real cosmic variance
caused by large-scale structure in the \lya\ forest.

\begin{figure}
\epsscale{1.2}
\plotone{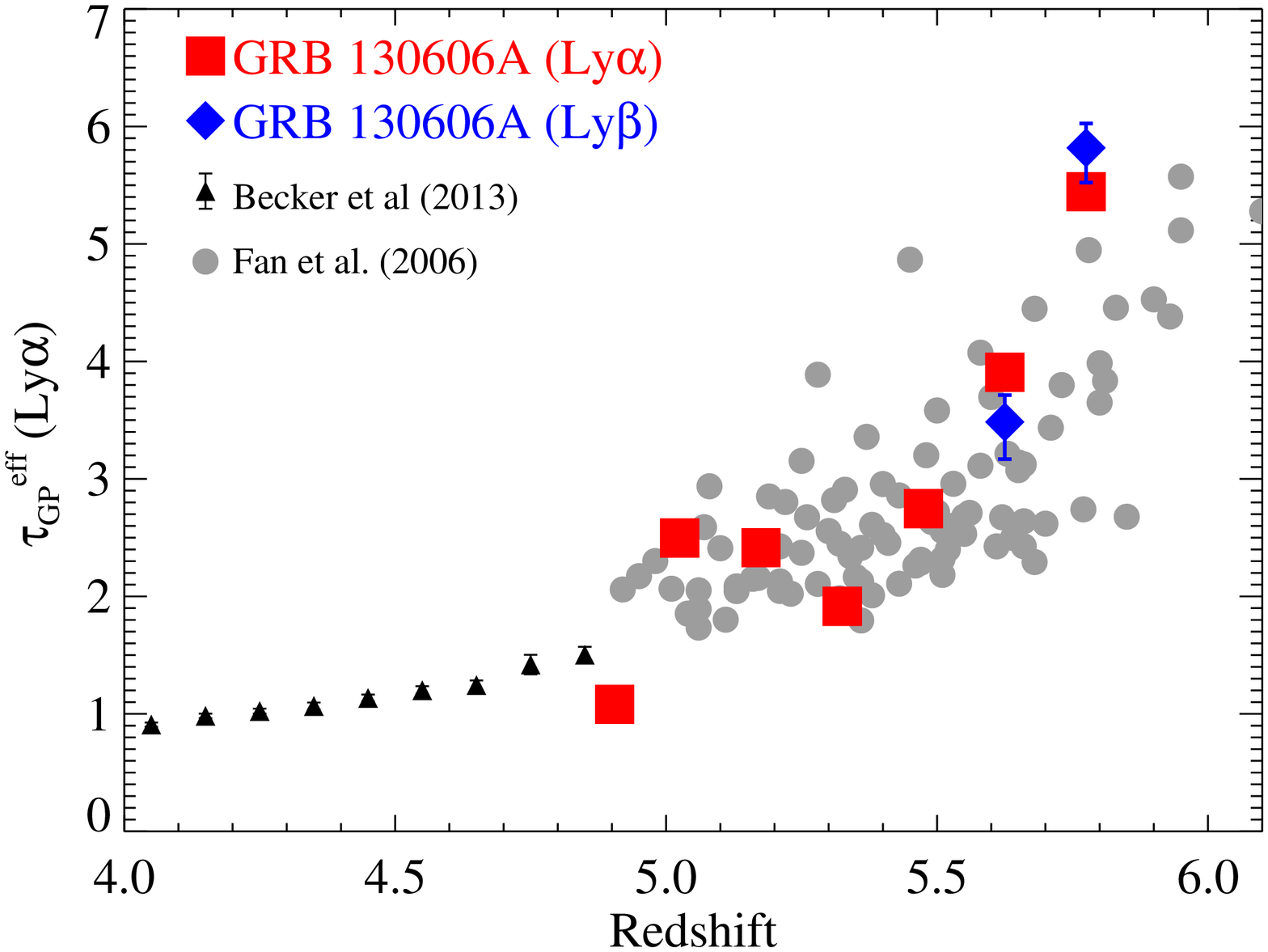}
\caption{The effective optical depth to \lya, \taueff(\lya),
  computed in 
  bins of width $\Delta z=0.15$ in both \lya\ and \lyb.  Our
  estimate of \taueff\ from the \lyg\ absorption in the
  highest-redshift bin ($z$=5.775) is 12.5 and thus off the scale of
  the rest of the points on the plot.  The error bars were computed
  from the extreme ``Low'' and ``High'' continuum models discussed in
  section 4.1 and represent bounds on the systematic uncertainties.
  The statistical errors are smaller than the plotted data points.
  The comparison points were measured in \lya\ 
  absorption of individual high-redshift quasars from \citet{fan06} and
  a large sample of lower-redshift quasars from \citet{becker13}.
}
\label{taufig}
\end{figure}

We convert our transmission measurements into an effective
optical depth following the definition of \citet{fan06} that
\begin{equation}
\taueff = -\ln(\mathcal{T}) ,
\end{equation}
where $\mathcal{T}$ is the average transmission relative to the
continuum.  This is only an effective rather than true optical depth
because transmission in a clumpy IGM with variable density and
ionization is dominated by low-density regions \citep{sc02,fan02,oh}.
We compute this \taueff(\lya) 
in bins of size $\Delta z$=0.15 relative to \lya\ to facilitate direct
comparison with the compilation of results from high-redshift quasars
of \citet{fan06}.  In
addition, we compute the same statistic for \lyg\ and \lyb\ in one and
two bins, respectively.  We use the same statistical correction as
\citet{fan06} for foreground \lya\ absorption and the same conversion
factors to determine \taueff(\lya) from \taueff(\lyb) and
\taueff(\lyg), so our results are calculated as consistently as
possible with the quasar data. We show the points from \grb\ in
Figure~\ref{taufig} and report the numbers in Table~\ref{transtab}.

The data points from the \grb\ line of sight are generally consistent
with the evolution of \taueff(\lya) seen along the quasar sight lines
and fall within the observed range of variation, although our highest
redshift bin has an optical depth on the high side relative to the
quasar measurements. Our single \lyg\ point,
covering $z$=5.70$-$5.85, is the first using that tracer at $z<6$ and
implies a substantially higher \taueff(\lya)=12.5 than that measured
from \lya\ or \lyb.  This was also seen by \citet{fan06} at $z>6$ in a
very limited number of sightlines.  \lyg\ measurements in the quasars
are made difficult by the necessity of avoiding Ly$\delta$ at the blue
end of the spectral window and the quasar's proximity zone at the red
end.  The interpretation of differences in \taueff\ determined from
multiple proxies is complicated because the differing strengths
of the transitions makes them differently sensitive to
inhomogeneity in the IGM and requires comparison to numerical
simulations.  Recent work has focused on the statistics of dark pixels
in the \lya\ forest as a probe of the neutral fraction in the
high-redshift universe to avoid dependence on models \citep{mcgreer}.
\citet{fan06} emphasized that the quasar data at $z>5.5$ show an
acceleration in the evolution of the effective optical depth with
redshift relative to a power law extrapolation from data at lower
redshifts \citep{song04,becker13}, and that the scatter increases with
redshift, possibly indicating that the tail end of reionization was
a patchy process.  The interpretation that there is an observed change
in slope of \taueff\ associated with late reionization has been
challenged by other studies (e.g., \citealt{becker07}).

\begin{deluxetable*}{lccccc}
\tabletypesize{\scriptsize}
\tablecaption{\grb\ \lya\ Transmission}
\tablehead{
\colhead{Redshift Range} &
\colhead{Line} &
\colhead{Spectrum\tablenotemark{a}} &
\colhead{Continuum\tablenotemark{b}} &
\colhead{Transmission} &
\colhead{\taueff (\lya)}
}
\startdata
4.86$-$4.95\tablenotemark{c} & \lya\ & G & BF & 0.406 &  0.90 \\
4.95$-$5.10 & \lya\ & G & BF & 0.089 &  2.41 \\
4.95$-$5.10 & \lya\ & G & L & 0.099 &  2.31 \\
4.95$-$5.10 & \lya\ & G & H & 0.084 &  2.48 \\
5.10$-$5.25 & \lya\ & G & BF & 0.100 &  2.30 \\
5.13$-$5.25\tablenotemark{d} & \lya\ & M & BF & 0.109 &  2.22 \\
5.25$-$5.40 & \lya\ & G & BF & 0.168 &  1.78 \\
5.25$-$5.40 & \lya\ & M & BF & 0.164 &  1.81 \\
5.40$-$5.55 & \lya\ & G & BF & 0.079 &  2.54 \\
5.40$-$5.55 & \lya\ & M & BF & 0.081 &  2.52 \\
5.55$-$5.70 & \lya\ & G & BF & 0.022 &  3.80 \\
5.55$-$5.70 & \lya\ & M & BF & 0.021 &  3.88 \\
5.55$-$5.70 & \lyb\ & G & BF & 0.228 &  3.33 \\
5.70$-$5.85 & \lya\ & G & BF & 0.005 &  5.29 \\
5.70$-$5.85 & \lya\ & G & L & 0.005 &  5.24 \\
5.70$-$5.85 & \lya\ & G & H & 0.005 &  5.29 \\
5.70$-$5.85 & \lya\ & M & BF & 0.016 &  4.13 \\
5.70$-$5.85 & \lyb\ & G & BF & 0.074 &  5.85 \\
5.70$-$5.85 & \lyg\ & G & BF & 0.059 &  12.5 \\
\cutinhead{Darkest trough}
5.71$-$5.83 & \lya\ & G & BF & $\lesssim$0.0017 &  $\gtrsim$6.36 \\
5.71$-$5.83 & \lyb\ & G & BF & 0.062 & 6.22 \\
5.71$-$5.83 & \lyg\ & G & BF & 0.026 & 13.0 \\
5.725$-$5.79 & \lya\ & G & BF & $\lesssim$0.0022 &  $\gtrsim$6.13 \\
5.725$-$5.79 & \lyb\ & G & BF & 0.019 & 8.86 \\
5.725$-$5.79 & \lyg\ & G & BF & 0.023 & 16.6 \\
\enddata
\tablenotetext{a}{G=Gemini, M=MMT}
\tablenotetext{b}{Continuum normalization: BF= Best fit, L = Low, H=High}
\tablenotetext{c}{Lower redshift limit truncated to avoid
  \lyb\ absorption from host}
\tablenotetext{d}{Lower redshift limit set by spectral range}
\label{transtab}
\end{deluxetable*}

\subsection{Uncertainties}

We now consider the sources of uncertainty in these optical depth
measurements.  We do not list formal error bars in
Table~\ref{transtab} because the statistical errors are completely
negligible and dominated by systematics.  To demonstrate this point,
we note that the median error bar per 1.38~\AA\ spectral pixel in the
GMOS data between $z_{\mathrm{Ly}\alpha}$ of 5 and 6 is 0.8\% of the
unabsorbed continuum.  The large redshift bins we use to compare to
the results of \citet{song04} and \citet{fan06} then average over many
such pixels ($\Delta z$=0.15 is $\sim$132 pixels).  Another check on
our errors is to compare our two independent spectra from different
instruments. Results from both MMT
and GMOS are given in Table~\ref{transtab} and are mostly 
similar, except for the darkest part of the \lya\ absorption between
redshifts 5.7 and 5.8, where the MMT data show more transmission.
This is caused by the extra bit of flux visible near
8150~\AA\ in Figure~\ref{specfig}.  This effect was caused by
inadequate flat fielding and removal of the slit function in the MMT
data leading to uneven background levels along the slit.
It results in a slight excess of flux per pixel after background
subtraction that is only visible near zero flux levels when the data
are binned up.  The exact amount of this positive flux was found to
vary  depending the exact background apertures chosen.  We
trust the GMOS data much more due to their 
significantly higher S/N and our ability to check with the
nod-and-shuffle reduction, which has much better control over the
flat-fielding and sky subtraction uncertainties at flux levels near
zero.

Our results are presented as transmission fractions, so we also need
to examine our assumptions about the proper level of the unabsorbed
continuum.  We have been using our best-fit power law so far in this
analysis.  In section 2, we noted that the continuum did appear to be
curved relative to a single power law.  Multiband photometry extending
to the NIR reported by the RATIR and GROND collaborations
\citep{ratirgcn,grondgcn} implies a bluer 
power law than the flatter continuum we have fit to our data over a
more limited wavelength range.  These observations can easily be
reconciled by a small amount of dust producing curvature in the
spectral energy distribution given that our observations are
at rest-frame wavelengths $<$1450~\AA.  However, we will demonstrate
that our results are robust for any reasonable value for the true
shape of the continuum.

We show two extreme hypothetical alternative continuum shapes in
Figure~\ref{specfig} as dashed lines.  All assumed continuum shapes
have to be constrained to pass near the actual observed continuum near
8650~\AA.  In addition, there is a peak in the IGM transmission near
7150~\AA\ that the true unabsorbed continuum must pass above, which
sets a limit on how red the continuum can be.  This power-law slope is
shown by the lower dashed line on the figure.  As a maximally blue
model, we take a power law with a slope as different from the best fit
model as the maximally red one, but in the opposite direction from the
best fit, and also constrain it to match the data near 8650~\AA.  We
normalize the GMOS data by both of these extreme models and computed
the transmission fraction in the same redshift bins as above.  We list
the results with these alternative continuum normalizations (called
``Low'' and ``High'') for two redshift bins in Table~\ref{transtab}.
Even in our bluest bins, at $z\approx5$, the difference in
\taueff(\lya) is only $\sim$0.1.  At higher redshift, closer in
wavelength to where the continua are normalized, the effect is even
smaller.  In part, this is  because \taueff\ depends only
logarithmically on the continuum normalization.  We emphasize that
these alternative normalizations are far larger than anything
motivated by the data and yet they do not materially affect the
results.  The maximal effects on the \grb\ data points in
Figures~\ref{lyafig} and \ref{taufig} induced by these choices of
continuum slope are smaller than the points on the plots.

We conclude that the uncertainties in our optical depth measurements
are negligible compared to the dominant systematics in interpretation
caused by theoretical uncertainties and cosmic variance along
different lines of sight.  In particular, these measurements would
benefit from an improved theoretical understanding of the differences
between \taueff(\lya) measured from the different Lyman series lines.

\subsection{Dark GP Trough}

We examine our spectra for continuous regions of extremely high
opacity and find that there is no detectable \lya\ transmission in the
redshift range of $z$=5.71$-$5.83, with a 3$\sigma$ upper limit of
$\mathcal{T}_{\mathrm{Ly}\alpha}\lesssim$0.2\%, or
\taueff(\lya)$>$6.4.  This is comparable in width and optical depth
to the lowest-redshift Gunn-Peterson \citep{gp65} troughs previously
claimed, which were identified by \citet{fan06} in the quasars SDSS
J104845.05+463718.3 \citep{fan03} and SDSS J125051.93+313021.9
\citep{fan_06a}.  However, there are clear spikes of
\lyb\ transmission at the ends of this redshift interval, including
the interesting peak near $z$=5.803 noted previously to be almost
coincident in redshift with a foreground absorber, so we also define a
narrower redshift range of $z$=5.725$-$5.79 to isolate the darkest
part of the trough. This more restricted interval is the one marked
with a black bar on Figure~\ref{transfig}.  The $\mathcal{T}$ and
\taueff(\lya) measurements for these two interval definitions are
tabulated in Table~\ref{transtab}. 

Despite the lack of detectable \lya, some flux is present in the
\lyb\ and \lyg\ windows, indicating that the redshift interval is far
from opaque and still highly ionized.  The transmission of
\lyg\ over the core redshift range implies that \taueff(\lya) is
$\sim$17.

\subsection{Neutral fraction in the IGM}

The afterglow spectra of GRBs can also be used to probe the neutral
fraction of the IGM \citep{jordi,bl04}.  If a high-redshift GRB occurs
when the universe still contains a substantial fraction of neutral
hydrogen, the red damping wing of this material will affect the shape
of the cutoff in flux at \lya. \citet{totani06} have searched for such
an effect in the spectrum of the $z$=6.295 GRB 050904 and found a best
fit consistent with zero neutral hydrogen, although their analysis was
hampered by the strong DLA of the host galaxy.  \citet{patel10} have
also performed an analysis on the $z$=6.733 GRB 080913 and again found
a null result. The much lower \lnh\ we determine here than for GRB
050904 and the higher S/N of our data relative to the spectrum of
GRB 080913 allow us a cleaner test, although the neutral fraction is
not expected to be sufficiently high to be detectable at this lower
redshift given the limits on \lya\ opacity discussed above.  

\begin{figure}
\epsscale{1.2}
\plotone{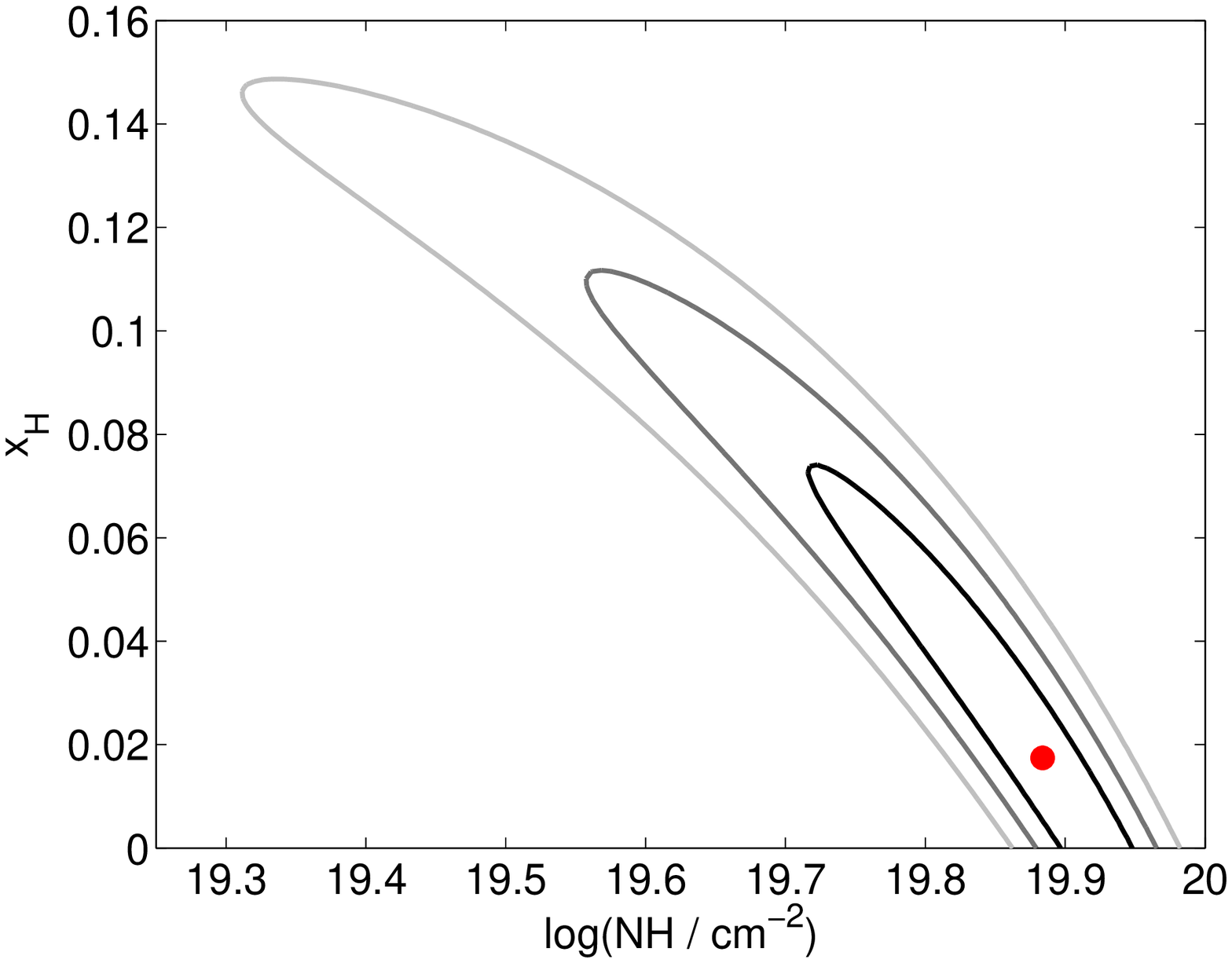}
\caption{Contours (1, 2, and 3$\sigma$) of fits to the red wing of
  \lya\ absorption when a simple model with constant $x_{\mathrm{H
    I}}$ in the IGM \citep{jordi} is allowed in addition to an
  absorber at the host redshift with a column density of \lnh.  
  Models treating IGM inhomogeneity more realistically will
  significantly relax these constraints \citep{mf08,mcquinn08}, but
  the best fit will still be consistent with zero neutral fraction.
}
\label{nhfig}
\end{figure}

We use the approximations of \citet{jordi} to model the IGM neutral
density as a constant over the redshift range of interest.  A pure IGM
fit to the \lya\ cutoff in the GMOS 
spectrum is a significantly worse fit than the single absorber model
from Section 3.  In addition, minima in the \lyb\ and
\lyg\ absorption spectra at the host redshift (Figure~\ref{transfig})
demonstrate the need for at least some absorption from the host
galaxy.  A combined fit with a host galaxy absorber (fixed to  
the redshift of the metal lines) along with the IGM model can fit the
data as long as the IGM neutral fraction ($x_{\mathrm{H I}}$) is below
$0.05$, but is not required by the data.  We conclude that our
spectra are consistent with zero neutral fraction and
$x_{\mathrm{H I}}<0.11$ at the 2$\sigma$ level (Figure~\ref{nhfig}). 
However, allowing more realistic models than a simple constant neutral
density in the IGM makes the interpretation of \lya\ damping wings
more problematic and significantly relaxes these constraints
\citep{mf08,mcquinn08}.

We note that although most GRB host galaxies are DLAs \citep{palli06},
two out of the three highest redshift events with good measurements of
the host hydrogen columns have \lnh$<20$ (GRBs 130606A
and 080913), which may bode well for the detectability of the red
damping wing effect in the future at higher redshift.  An evolution in
the \lnh\ values observed in GRB afterglow spectra would also have
important implications for reionization.
\citet{chen07} have used the distribution of \lnh\ values observed in
GRB afterglow spectra as a proxy to measure the escape fraction of UV
ionizing photons from star-forming galaxies.  If the GRB host
galaxy \lnh\ distribution really is lower at higher redshift, then
the higher escape fractions would imply that star-forming
galaxies can more efficiently reionize the universe.  Data from more
lines of sight at these high redshifts are necessary to test this
hypothesis. 

\section{Conclusions}

We have presented high S/N spectra of the optical afterglow of
\grb\ at $z$=5.9134, the first high-redshift afterglow to have a
dataset of similar quality for IGM studies to those published for
individual high-$z$ quasars, although our spectral resolution was not
as high as in the best quasar datasets (e.g., \citealt{white03}). 
For comparison, we estimate that the continuum magnitude at the time
of the GMOS observations was $\sim$19.6 mag \citep{perley,grondgcn}, or
$M_{\mathrm{1250\AA}}\approx-27$~mag (AB).  This is comparable to the most
luminous quasars known at similar redshifts \citep{fan06} and we were
able to obtain 1.3~hr of spectroscopy with the 6.5-m MMT and 2~hr with
the 8-m Gemini-N telescope. 

These observations represent the first dataset on the evolution of the
IGM opacity at these redshifts using a tracer other than quasars,
which \cite{mes10} has argued are sufficiently biased tracers of
large-scale structure that they will overestimate the degree of
ionization of the IGM.  It is therefore reassuring that the general
trend of the quasar observations is reproduced in our dataset,
although an individual sightline is of limited utility for making firm
conclusions because of cosmic variance.  Once we have obtained a
sample of GRB afterglow spectra at high redshift, it will be
interesting to compare the statistics of \lya\ absorption using both
tracers.  

We also find an extended region of \lya\ absorption from $z$=5.71 to
5.83, similar to the lowest-redshift Gunn-Peterson troughs found in
quasar absorption spectra, over which we place a 3$\sigma$ upper limit
of 0.2\% on the \lya\ transmitted fraction, although \lyb\ and
\lyg\ are not completely absorbed.  The pixel-scale statistics (e.g.,
\citealt{mcgreer}) of dark regions in \lya\ absorption windows such as
this in a larger sample of GRB afterglows will offer a complementary
view of reionization to the studies of quasars.

In addition, we have identified numerous metal absorptions on the
bright GRB afterglow continuum at wavelengths redward of \lya\ at the
host redshift, due to both the IGM 
and the ISM of the host galaxy.  A metal absorption system at
$z$=5.806 appears to be correlated with a region of slightly enhanced
transmission in the \lyb\ and \lyg\ forests.  We have used the host
ISM absorption lines to bracket the gas phase abundances for this
star-forming galaxy at $z$=5.913 between [Si/H]$\gtrsim$$-1.7$ and
[S/H]$\lesssim$$-0.5$.  The low hydrogen column density in this host
galaxy (\lnh = 19.93$\pm$0.07) as well as that of the $z$=6.7 GRB
080913 may be evidence for an evolving escape fraction for UV photons
from star-forming galaxies at high redshift.

\acknowledgments
We thank the Gemini and MMT staffs for their assistance in obtaining
these observations. 
The Berger GRB group at Harvard is supported by the National Science
Foundation under Grant AST-1107973 and by NASA/{\it Swift} AO8 grant
NNX13AJ64G. 
Based in part on observations obtained under Program ID
GN-2013A-Q-39 (PI: Cucchiara) at the Gemini Observatory, which is
operated by the Association of Universities for  
    Research in Astronomy, Inc., under a cooperative agreement with the
    NSF on behalf of the Gemini partnership: the National Science
    Foundation (United States), the Science and Technology Facilities
    Council (United Kingdom), the National Research Council (Canada),
    CONICYT (Chile), the Australian Research Council (Australia),
    Minist\'{e}rio da Ci\^{e}ncia, Tecnologia e Inova\c{c}\~{a}o (Brazil)
    and Ministerio de Ciencia, Tecnolog\'{i}a e Innovaci\'{o}n Productiva
    (Argentina).  
Some observations reported here were obtained at the MMT Observatory,
a joint facility of the Smithsonian Institution and the University of
Arizona.  

{\it Facilities:} \facility{Gemini:Gillett (GMOS-N)}, \facility{MMT
  (Blue Channel Spectrograph)}

\end{document}